\documentstyle[12pt,aaspp4,amssym]{article}
\input psfig.sty
\newcommand{\zphot}{$z_{phot}$ }
\newcommand{\zspec}{$z_{spec}$ }
\begin{document}

   \title{High-Redshift Clustering in the HDF}

   \author{J.M. Miralles
           \and R. Pell\'o}

   \authoraddr{Observatoire Midi-Pyr\'en\'ees, 14 Avenue E. Belin, F-31400 Toulouse, France}

   \authoremail{miralles@obs-mip.fr, roser@obs-mip.fr}

   \affil{Observatoire Midi-Pyr\'en\'ees, Laboratoire d'Astrophysique de Toulouse UMR 5572,
 14 Avenue E. Belin, F-31400 Toulouse, France}

\received{1997} 
\accepted{2019}

\righthead{High-Redshift Clustering in the HDF}

\begin{abstract}

This paper addresses the problem of detecting high-redshift clustering 
in deep photometric surveys. We have used photometric redshifts to 
select different samples of galaxies in the HDF, in order to study their
clustering properties. Errors and biases associated with the photometric 
redshift techniques have been carefully studied through simulations as a 
function of the photometric uncertainties, the redshift domain and the 
galaxy type. A direct comparison between photometric and spectroscopic 
measures of the redshift was also performed. We have studied the 2-D 
distribution and clustering of galaxies within the $2.5 \lesssim z \lesssim 
4.5$ domain, in redshift bins of 0.5, using different techniques such as the
Kolmogorov-Smirnov test, the galaxy-galaxy correlation function and the
number density counts. Although a clustering signal at small scales is
detected in the whole redshift interval, the strongest signal at large
scales is found in the redshift bins including z$\simeq$3.4. For these
bins, the clustering length is $r_0 = 0.16 \pm 0.03$ h$_{50}^{-1}$ Mpc,
leading to a present correlation length of $r_0 = 4.1 \pm 0.8$ h$_{50}^{-1}$
Mpc (q$_0$=0.1) assuming linear evolution. An excess appears in the correlation 
function of chip 2 with respect to the fit, at angular scales from 30" to 45". 
This excess could be associated to the main structure detected in this field, 
which contains $\sim 20\%$ of the objects identified at $3.4 \lesssim z \lesssim 
3.9$, among which 2 galaxies with spectroscopic z$\simeq$3.4.
 Its dimensions are 3 h$_{50}^{-1}$ Mpc $\times$ 0.5 
h$_{50}^{-1}$ Mpc (comoving coordinates, q$_0$=0.1). The galaxies at $3.4 
\lesssim z \lesssim 3.9$ exhibit a SFR of a few solar masses 
per year typically, but their comoving density is a factor of $\sim 50$ higher 
than the population of star-forming galaxies reported by Steidel et al. (1996b).
 The resulting star formation rate density is at least $1.1 \times 10^{-2}$ 
h$_{50}$ $M_{\odot}$ $yr^{-1}$ $Mpc^{-3}$ ($1.8 \times 10^{-2}$ 
h$_{50}$ $M_{\odot}$ $yr^{-1}$ $Mpc^{-3}$) with q$_0$=0.1(0.5), slightly 
higher than the results by Madau et al. (1996) at $2.5 \lesssim z \lesssim 3.5$, and 
then incompatible with a global decrease of the star formation in this 
redshift domain. These results on star formation and clustering are consistent 
with a hierarchical scenario for galaxy formation.

\end{abstract}


\section{Introduction}

The detection and the study of high-redshift clustering is one of the main deals in modern 
observational cosmology. The clustering parameters observed at very high-redshifts
constitute a discriminating test for the different cosmological models, and they can
greatly help on constraining the cosmological parameters (e.g. White 1996, Baugh et al 1997, 
Moscardini et al. 1997). During the last year, several models of galaxy formation 
in hierarchical scenarios have been proposed to account for the recently observed 
properties of the galaxy population (Baugh et al. 1997, Moscardini et al. 1997, 
Steidel et al. 1998). The emerging picture is that the Lyman-break galaxies observed 
at 2.5 $\lesssim z \lesssim$ 3.5 are strongly biased mass tracers, associated with large 
dark matter halos. Their clustering and star-formation characteristics are well reproduced
using CDM models, with a relatively high bias parameter ($b \sim 4$). The brightest 
galaxies in the sample by Steidel et al. (1996a,1996b) (hereafter ST96a and ST96b) 
should be interpreted as the progenitors of the brightest present-day galaxies, 
preferentially located in clusters or groups of galaxies. In order to constrain 
the theoretical models, it is crucial to measure the clustering properties 
(through the correlation length) versus the star formation rate density as a function 
of the redshift.

The problem for performing such measures comes from the extreme difficulty to identify the 
very faint population of galaxies and to assign a redshift to each object. Most of the time, 
galaxies at high redshifts are too faint to be studied spectroscopically, and 
sometimes even too faint to be identified in optical ground-based images.
Despite the difficulty, several evidences for high-redshift clustering have appeared recently, 
such as the groups or clusters of galaxies reported (increasing in redshift) at z=2.06
behind CL0939+47 (Dressler et al. 1993), at z=2.38 (Francis et al. 1996), at z=3.14 
(Le F\`evre et al. 1996) and at z=3.4 (Giavalisco, Steidel \& Szalay 1994). The first
spectroscopic and statistically significant samples of high-redshift galaxies are 
presently being built (ST96a, ST96b, Steidel et al. 1998). Steidel et al. (1998) have  
reported the discovery of a large scale structure at $z \sim 3$ based on these data.
Waiting for the extensive spectroscopic surveys to come in the near future, 
photometric redshift techniques provide with a useful tool to extend the present 
surveys up to the faintest magnitude limits, and for a larger number of galaxies
(Connolly et al. 1995, Sawicki, Lin \& Yee 1997 (SLY), Subbarao et al. 1996, among others).
They have already proved their value on identifying high-redshift galaxies (ST96a, ST96b) 
and also on unveiling clusters at moderate redshifts (Pell\'o et al 1996, 
Connolly et al. 1996). Nonetheless, the accuracy on the redshift, in this case, is strongly 
dependent on the photometric accuracy.

The Hubble Deep Field (Williams et al. 1996) offers the necessary deepness, spatial resolution 
and photometric accuracy to search for structures, especially at z $\geq$ 2. Villumsen, 
Freudling \& Da Costa (1997) have studied the angular correlation function for galaxies in 
the HDF, w($\theta$), and, according to them, the clustering signal can be measured up to 
R=29. Their results are consistent with a linear evolution of the clustering, giving a 
present-day correlation length of r$_0 \sim$ 4 h$^{-1}$ Mpc, the same as observed for 
IRAS galaxies (Fisher et al. 1994). Besides, they do not detect as expected the effects of 
the magnification bias. Their results evidence the absolute need for a redshift estimate to 
conclude about the clustering properties at high-redshift. The situation is better at lower 
redshift (z $<$ 1.5), where Cohen et al. (1996) have studied the clustering by means of a 
large spectroscopic survey on the HDF and the surrounding fields. They found some evidence
for clustering in the redshift space, in particular two peaks at redshifts 0.5 and 0.8,
but they have been unable to detect any particular spatial structure to which 
these peaks could be associated.
 
In this paper we investigate the existence of structures at high-redshift (z $\geq$ 2.5) in 
the HDF, using photometric redshift techniques to select the different populations of galaxies.
The aim is to access the redshift information for a sample of faint galaxies as large as possible, 
and to perform a combined study of their clustering versus spectrophotometrical properties
as a function of the redshift. A similar technique has been applied to the HDF by other authors
aiming to study the properties of the faint population of galaxies (see for instance 
Sawicki et al. 1997; Lanzetta, Yahil \& Fernandez-Soto 1996 (LYF)), and in general a reasonable 
match was noticed between spectroscopic and photometric redshifts, even at high-redshift.

In \S 2 we describe the method used to compute photometric redshifts in the HDF.
The accuracy of the results and the possible biases are discussed on the basis of simulations, 
especially at the magnitude levels of the objects expected at high-redshift. A
direct comparison between spectroscopic and photometric redshifts for low-z and high-z galaxies is also 
given. The photometric redshift distribution derived for this field is briefly presented in \S 3.
We also study in this section the detailed 2-D distribution of the selected galaxy samples at 
high-redshift, and we analyze the reliability of the results using different tests for clustering, 
including the spatial correlation function. The main photometric characteristics  
of this population of high-redshift galaxies are highlighted in \S 4, where we compute 
the star formation rate density for the strongly clustered population. Finally, we discuss 
the results in \S 5, as well as the constraints and implications for cosmological models
that can be derived from them, including the future perspectives of this work.
Unless otherwise stated, we use $H_0=50 km s^{-1} Mpc^{-1}$ and $q_0=0.1$.

\section{Photometric redshifts}

\subsection{The method}

   The technique used to compute photometric redshifts
(hereafter \zphot) is a standard $\chi^2$ minimization procedure.
The observed spectral energy distribution (SED) of each galaxy, as obtained
from its multicolor UBRI photometry, is
compared to a set of template spectra. The aim is to find the best matching 
\zphot which minimizes the $\chi^2$, defined as:

\begin{equation}
 \chi^2=\sum_{i = 1}^{N_{filters}} \left( {F_i^{O}-F_i^{T}\over \sigma(F_i)} \right) ^2
\label{chi2}
\end{equation}

where $F_i^O$, $F_i^T$ and $\sigma(F_i)$ are, respectively, the observed and the 
template fluxes in the i band, and the uncertainty associated to the 
photometric errors in the same filter. $F_i^T$
is normalized to match the observed flux in an arbitrary reference band.
The four filters are F300W (U), F450W (B), F606W (R) and F814W (I)
(see Williams et al 1996, and the WFPC2 Instrument Handbook 1995, from 
which the filter responses were taken). 
The difference with respect to other similar methods (Gwyn \& Hartwick
1996, Sawicki et al. 1997 among others) is the large number of
template spectra used here. The new Bruzual \& Charlot evolutionary
code (GISSEL96, Bruzual \& Charlot 1993, 1997) has been used to build 5 
different synthetic star formation histories, roughly matching the observed 
properties of local field galaxies: a pure
burst of 0.1 Gyr, a constant star-forming system, and three $\mu$ models
(e-decaying SFR) with characteristic time-decays chosen to match
the sequence of colors for E, Sa and Sc galaxies. For each of these types,
we select 51 synthetic spectra corresponding to different relevant
ages for the stellar populations, in order to closely follow all the
significant changes in the theoretical SEDs. 

   The template database includes 255 synthetic spectra in total. Nevertheless, 
the effects of metallicity or ISM (in particular, the presence of dust or emission 
lines) have not been included. According to our simulations (next section), such effects 
are of second order compared to the main sources of signal with a spectral resolution 
of $\sim$ 1000 \AA: the Lyman dropout and the Balmer or the 4000 \AA \ breaks, which 
are the most important features in the UBRI spectra of galaxies up to $z \sim 5$.

   The photometric catalogue for the whole field was obtained through the
SExtractor package (Bertin \& Arnouts 1996). The detection of objects on the
different images was made at $2 \sigma$ level above the sky background, with a 
minimum size requirement of 5 contiguous pixels (1 pixel=0.1") above this 
detection limit. The typical size of the faintest objects detected is 10 
to 15 pixels at $1 \sigma$.
Table \ref{cat} summarizes the properties of the catalogue. 1588
objects were detected on the whole field, at least in B, R and I filters, and
the total number per chip is the same within the statistical noise. Each multiple
object showing several bright regions less than 5 pixels apart 
(within a $10 \times 10$ pixels window) was 
considered as a single object, and its magnitudes and colors were obtained
through the integration of the fluxes within the whole region. Errors derived
from SExtractor are given in Table \ref{err1}.

\subsection{Estimate of errors and biases through simulations}

We have studied through simulations the accuracy of \zphot as a function of
the relevant parameters, namely the photometric errors, the filter
bands available and the galaxy type. For each test galaxy, the
difference between the \zphot and the model $z$ has been computed,
as well as an estimate of the individual uncertainty, $\Delta z$,
defined as:
\begin{equation}
 \Delta z =  0.5 \times [ z(+75 \%) - z(-75 \%) ]
\end{equation} 
where $z(+75 \%)$ and $z(-75 \%)$ are the \zphot limits to a
75 \% confidence level derived from the $\chi^2$ value. Different sets of simulated
catalogues were created for this exercise through the GISSEL96 
library, with galaxies distributed in redshift between 0 and 5. Photometric errors
were introduced as gaussian noise distributions of fixed FWHM for
each HDF filter band, and uncorrelated for different filters.
The first catalogue includes 800 galaxies, basically reproducing the
photometric properties of two extreme spectrophotometric types of
galaxies with solar metallicity, taken at different ages, with and without
evolution: : E/S0 galaxy (evolving 0.1 Gyr burst, $z_{form} = 5.3$,
aged 15 Gyr at z=0, with $H_0=50$ $km$ $s^{-1}$ $Mpc^{-1}$
and $q_0=0.1$), and a constant star forming system. Additionally, 
we have produced a second set of catalogues in order to compute
the mean accuracy of \zphot as a function of the photometric errors only,
each one containing $\sim 2000$ galaxies
uniformly distributed between $z =0$ and $5$, with randomly assigned types between 1 and 8.
 These 8 types correspond to the same 5 given above plus
3 additional e-decaying SFRs chosen to match S0, Sb and Sd types. 

   The results of the whole simulations are summarized 
in Table \ref{sim1}, where the following information for each 
simulated set of galaxies is given: the standard deviation of the
differences between \zphot and the model $z$ ($\sigma z$ (z)),
the mean systematic bias ($\Delta z = z(model) - $ \zphot), 
the mean individual uncertainties at 75 \% confidence level,
and the percentage of catastrophic identifications. We exclude the
catastrophic identifications ($ \mid z(model) - $ \zphot  $
\mid \geq 1.0 $) when computing these values. 
The results corresponding to the second catalogue are labeled as "all".
The dispersion is weakly dependent on the galaxy type, provided that the
evolving population of stars is older than
$\sim 10^7$ yr typically, a limitation arising from the ages of the stellar 
populations used to build our templates. It is worth noting that
no contribution from the ISM of galaxies has been taken into account
in these simulations, neither on the template spectra nor in the
test SEDs. In particular, the presence of emission lines 
in real SEDs has to be considered as noise included in the photometric 
uncertainties (0.1 to 0.2 magnitudes at worst). 
 
\begin{figure}
\centerline{\psfig{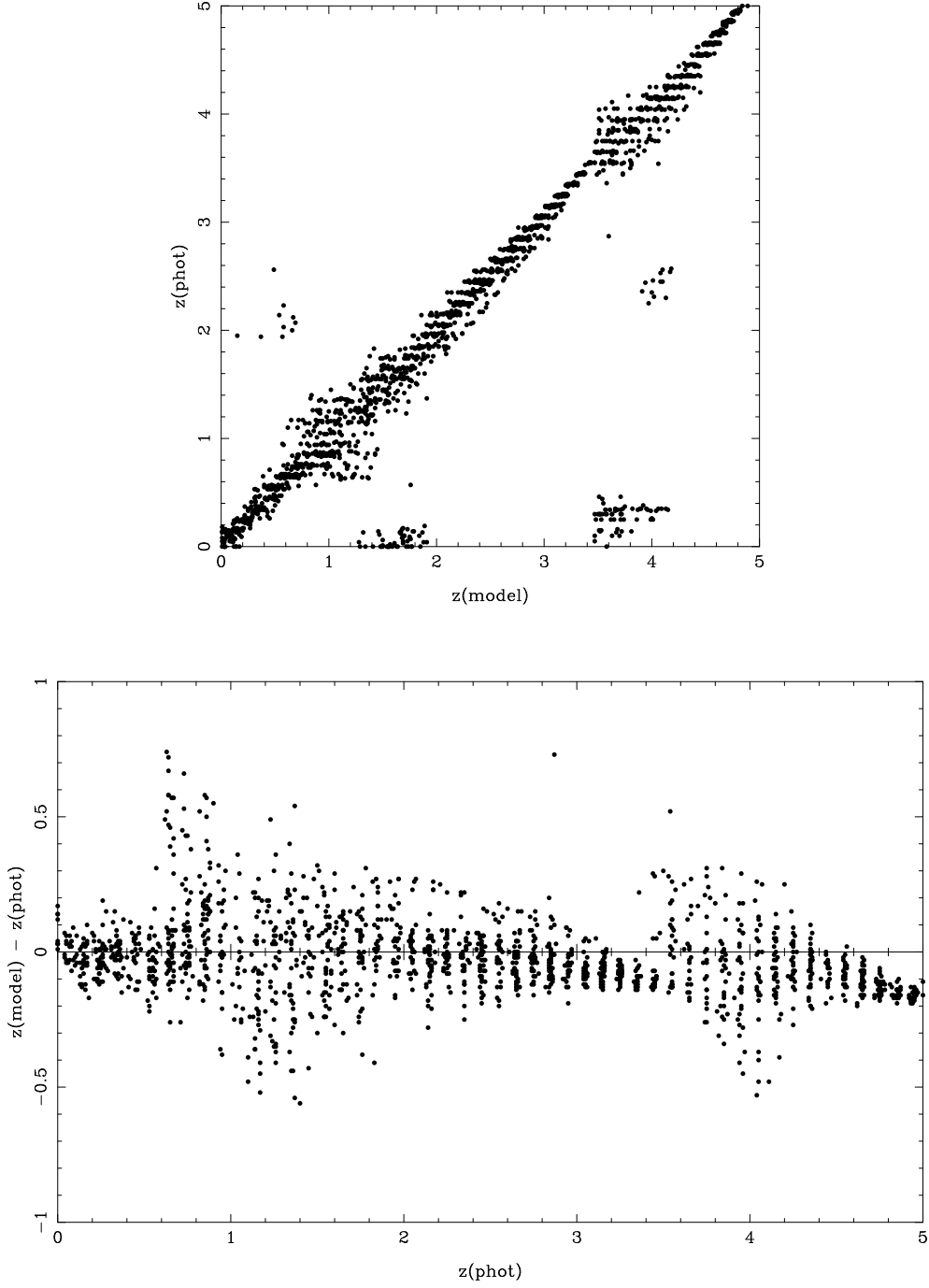}}
\caption{{\it Top panel:} \zphot vs model redshift (z(model)) for all the simulated galaxies 
with $\Delta m = 0.1$ in all the filters and solar metallicity. {\it Bottom panel:} 
z(model) - \zphot vs \zphot for the same sample, excluding the catastrophic regime
with $ \mid z(model) - $ \zphot  $ \mid \geq 1 $.}
\label{z_p}
\end{figure}

   The dispersion in the estimate of \zphot is
strongly dependent on the photometric uncertainties.
There is no significant gain for $\Delta m \lesssim 0.1$, but 
the dispersion and the number of multiple solutions with similar weight 
increases quickly up to $\Delta m \sim 0.3$, which is probably a limiting 
value for individual objects, and worse than the estimated errors in the HDF
(Table \ref{err1}). Figure \ref{z_p} displays the result of these
simulations when $\Delta m = 0.1$.
About $85 \%$ of the simulated objects have \zphot 
determined over $75 \%$ confidence level up to $\Delta m \sim 0.1$.
For this photometric accuracy, the mean number of catastrophic identifications, 
all types joined together, ranges between 1 and 20 \% typically. Nevertheless, 
there are 2 redshift domains with a poorly determined \zphot. The first one
is the $ z \lesssim 0.4$, with about 30 \% of catastrophic or multiple 
identifications when $\Delta m \ge 0.2$. The deviant objects are among the bluest in our
simulations, so a systematic bias exists against these particular SEDs
at low redshift. The second one is the $0.7 \lesssim z \lesssim 1.8$
domain, which is more noisy, as expected because of the lack of strong spectral
features going through the filter bands used for the HDF. More precisely, the
$1.4 \lesssim z \lesssim 1.8$ domain has to be avoided because the 
uncertainties on individual galaxies become huge and the number of 
degenerate solutions reaches 65 \%. We have performed a reduced number of
simulations including the near-IR J and K photometry. The results show as
expected that adding the J and K filters allows to solve the degeneracy
problem in this particular redshift interval,
without improving significantly the accuracy of \zphot outside it.

All the simulated galaxies with $z \ge 3.4$ are faded in U, with the 912 \AA\
break within the B filter. The results obtained in this particular redshift 
interval include this effect. It is important to note that all the catastrophic
identifications in the $3.4 \lesssim z \lesssim 4.5$ domain tend to underestimate
the population in this interval, without a significant contribution in the opposite
sense (low-redshift objects misidentified as high-redshift ones). This effect
can be appreciated in figure \ref{z_p}. 

We have also computed a specific set of simulations
to study the effects of metallicity and reddening on the results, especially 
at high redshift. For this purpose, a simulated catalogue was built, 
containing 1500 objects uniformly distributed between $z=0$ and $5$, with 
randomly assigned metallicities and reddening values. The reddening was taken between 
0 and 3 magnitudes in the V rest-frame, according to the SMC extinction 
curve by Pr\'evot et al. (1984). The results are weakly dependent on the 
reddening. A more detailed simulation would be required, including
a careful modelling of the UV rest-frame SEDs, to discuss the reddening/metallicity 
biases on the \zphot in details. This point is out of the scope of the present 
work. Nevertheless, according to our results 
in a reduced set of spectroscopic data, neither extinction nor 
metallicity effects will change the above results significantly.

The dispersion in \zphot is quite similar to the values found in the literature, 
even when the techniques used are appreciably different (Brunner et al. 1997, 
Connolly et al. 1995), but it is hard to compare the 
accuracy of our \zphot results as a function of the relevant parameters (photometric 
errors in particular) to other similar works. 

\subsection{Photometric versus spectroscopic redshifts}

\subsubsection{Spectroscopic Sample at z$<$1.5}

\begin{figure}
\centerline{\psfig{figure=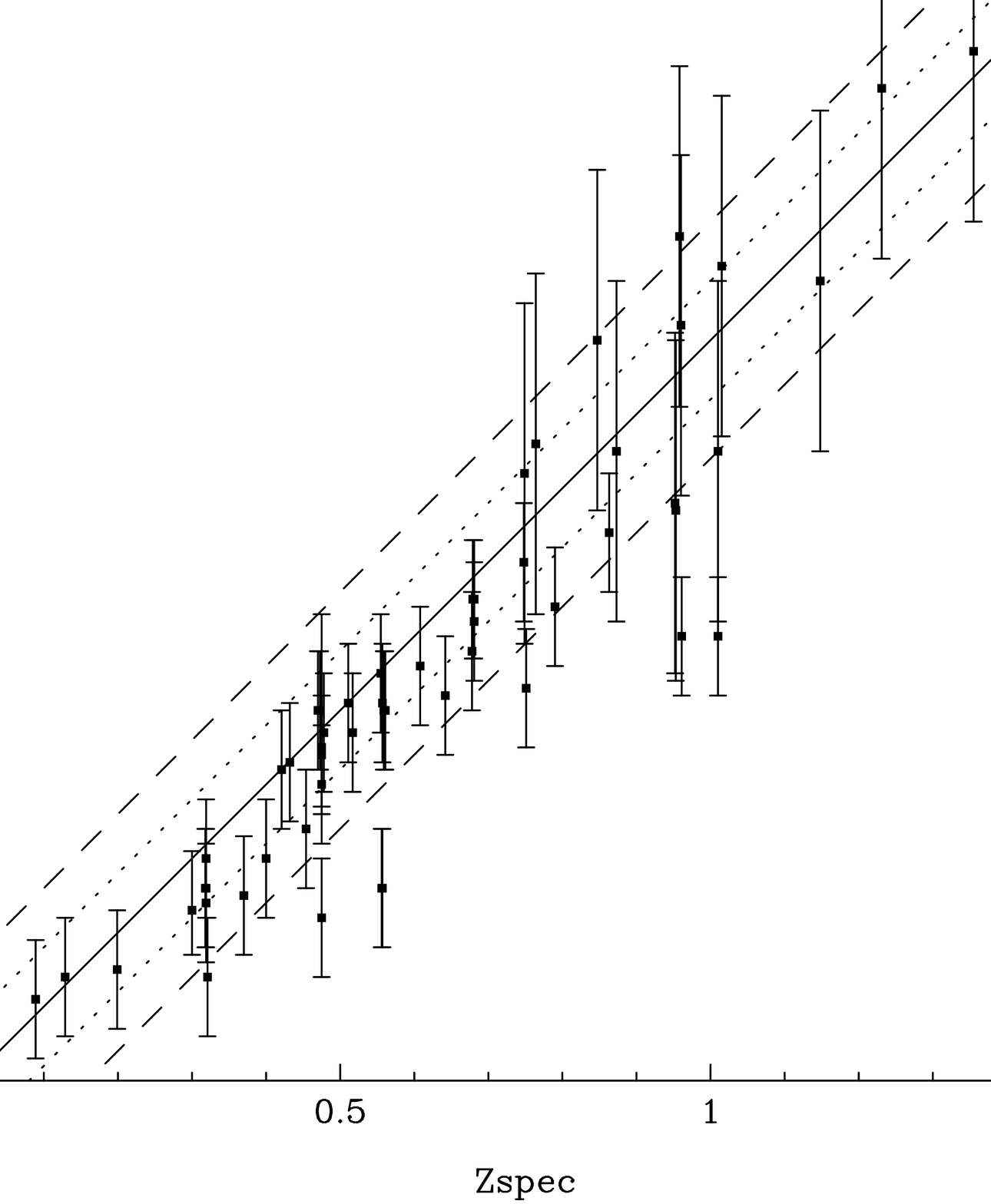,height=16cm}}
\caption{\zphot vs \zspec for the z$<$1.5 sample. Error bars represent
$\pm \sigma z$, the expected uncertainty according to simulations. Dotted and dashed lines 
correspond to the $\pm 0.08$ and $\pm 0.16$ error intervals, respectively.}
\label{zspec1}
\end{figure}

We have computed the \zphot for a sample of 52 objects observed by the Caltech,
Hawaii and Berkeley groups (respectively Cohen et al. 1996, Cowie 1997, Zepf et al. 1996), 
in order to check the accuracy of \zphot versus the spectroscopic 
redshift (\zspec). The highest redshift in this sample is 1.355, and all the 
galaxies have been detected in the four UBRI filters. 
Figure \ref{zspec1} shows the plot of \zphot versus \zspec for these objects, where the error
bars correspond to $\pm \sigma z$, according to the simulations.
When comparing the 2 values of the redshift, the standard deviation obtained is
$\sigma$=0.08, increasing to $\sigma \sim$ 0.15 towards z$\sim$1, in good agreement with the
above simulations. The \zphot computed remains in general quite close
to the observed spectroscopic redshift and no major systematic bias is observed. The 
average of (\zspec -- \zphot) is 0.05 for the whole sample, a value which is not highly
significant because it is smaller than the standard deviation. The maximum difference 
measured between \zphot and \zspec is 0.41, a value which is attained for a particular galaxy 
at \zspec=1.01 but found with a \zphot=0.6.
We have compared these results to two other similar works by Sawicki et al. (1997) 
and Cowie et al. (1997). The methods used are different but the statistical behaviours 
are very similar, even if some results for 
individual galaxies are in disagreement. The present dispersion is similar to
that of Sawicki et al. (1997) using the same sample of galaxies ($<$\zspec -- \zphot$>$=0.05 
and ${\sigma}_{SLY}=0.12$), despite the fact that these authors claimed that pure GISSEL 
models cannot be used for this exercise. The \zphot obtained by Cowie et al. (1997) 
over-estimates the redshift by 0.03 and their ${\sigma}$ is 0.1.

\subsubsection{Spectroscopic Sample at z$>$2.2}

We have also compared the \zphot and \zspec values for high-redshift
galaxies with secure redshifts in the range 2.2$<$z$<$3.5, taken from ST96a and 
Lowenthal et al. (1997) (hereafter LW97). This sample is limited to 
16 galaxies, most of them completely extinguished in the U band because the Lyman break moves
into the B band. For this reason, the photometry is only available in the B, R and I filters. 
To compute the \zphot for the U-dropouts, we fixed their U magnitude to 29.5, which 
represents 1$\sigma$ level over the sky background for the mean isophotal radius of the 
sample, and with a fixed error of 0.5.

\begin{figure}
\centerline{\psfig{figure=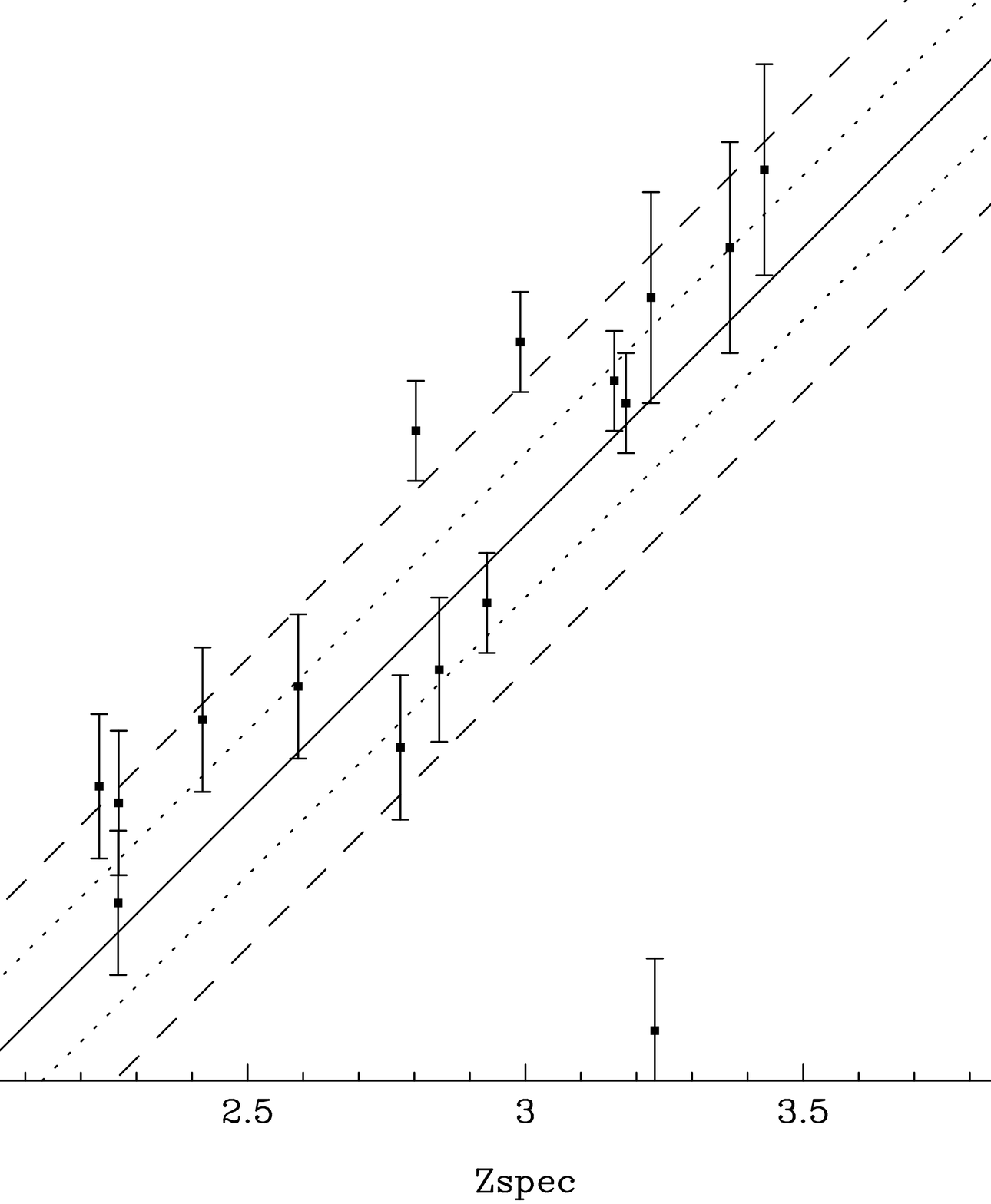,height=16cm}}
\caption{\zphot vs \zspec for the sample at z$>$ 2.2. Error bars represent
$\pm \sigma z$, the expected uncertainty according to simulations. Dotted and dashed lines 
correspond to the $\pm 0.13$ and $\pm 0.26$ error intervals, respectively.}
\label{zspec2}
\end{figure}

Results are shown in figure \ref{zspec2}.
The raw distribution shows an averaged scatter of 0.05 ($<$\zspec -- \zphot$>$ = -0.05) and a 
standard deviation $\sigma$=0.22. Compared to the z$<$1.5 galaxies, the dispersion is 
twice higher, but 
noticeably lower than the previous published values ($\sigma_{SLY}=0.28$, $\sigma_{LYF}=0.4$).
Removing from our sample a galaxy which is extremely discordant, with a $\Delta$z=1.14, 
the standard deviation reduces to $\sigma$=0.13, surprisingly close to the value expected 
from the above simulations. 
We are then confident that in this redshift range \zphot is a fair estimator of the redshift. 

\section{Results}

\subsection{Redshift Distribution of galaxies}

\begin{figure}
\centerline{\psfig{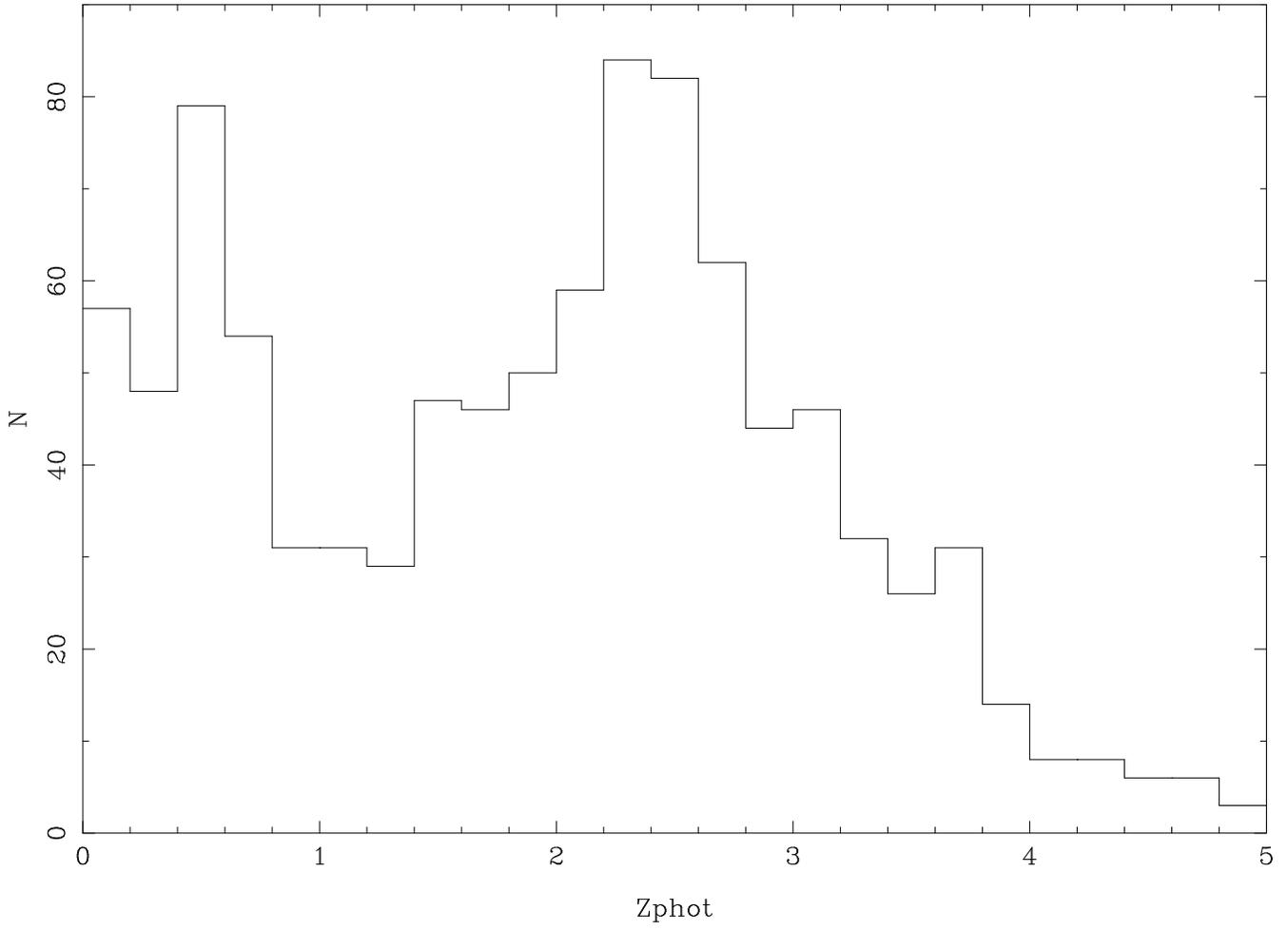}}
\caption{Photometric redshift distribution for galaxies detected 
at least in B, R and I filters, and with \zphot determined with a confidence level better than 75\%.}
\label{zdist}
\end{figure}

Among the population of objects detected at least in the three filters B, R and I, 1209 of 
them, corresponding to $76 \%$ of the total sample, have a \zphot value which is well 
determined above the 75\% 
confidence level. Figure \ref{zdist} shows the redshift distribution of these galaxies, where
$\sim$25\% of the sample lies at z$\leq$1 and $\sim$50\% is at z$\geq$2. The region
of 0.8$<$z$<$1.8 is extremely noisy in our simulations, and the redshift distribution is 
then less reliable.
Two main features appear: one peak at low redshift (0.4$\leq$z$<$0.6) and a second one at 
high-redshift (2.2$\leq$z$<$2.6). The former could be explained as an artifact. According 
to our simulations, up to $\sim$20-30\% of the objects in the 3.6$\leq$z$\leq$4 interval 
could be misidentified as low-z galaxies with z$\sim$0.5.
This happens when the Lyman break is mistaken for the 4000 \AA\ or the Balmer break. On 
the contrary, the distribution
in redshift is extremely reliable at \zphot$>$2, where the contamination by misidentified 
galaxies is expected 
to be small. A direct comparison with other similar works is difficult because the 
method and the selection criteria used are not the same.

\subsection{Clustering at z$\sim$3.4}

We have studied the 2-D distribution of photometrically identified high-redshift galaxies,
some of which have been already confirmed spectroscopically by ST96a and LW97.
A first approach has been to study each chip individually, except chip 1 (PC) where the 
detection level is lower. 
For each chip, we cut the catalogue in redshift slices of 0.5, going from z=2.5 to z=4.5. The 
thickness of one slice corresponds roughly to the worst dispersion estimated from the simulations.
To test whether the samples are consistent with a uniform distribution across the field,
a two-dimensional Kolmogorov-Smirnov (hereafter K-S) test (Peacock 1983) has been applied.
The results for each chip and redshift slice are summarized in Table \ref{ks}.
It appears that the 2-D distribution of galaxies in the interval [3.5;4] on chip 
2 has an extremely low probability, P$_{KS}$=0.4\%, to be drawn from a uniform distribution.
This chip contains 2 galaxies already identified at z$\sim$3.4 by LW97, plus 2 additional ones 
tentatively identified at the same redshift.
The \zphot of these 2 secure galaxies are \zphot=3.64$^{+0.15}_{-0.05}$ for galaxy A at \zspec=3.43 
and \zphot=3.50$^{+0.01}_{-0.01}$ for galaxy B at \zspec=3.368, the error bars corresponding 
here to the 90\% confidence level. For the 2 tentatively identified at \zspec=3.35 and \zspec=3.37, 
we obtained \zphot=3.56 and \zphot=3.5$^{+0.15}_{-0.12}$ respectively.
All these galaxies are then included in our [3.5;4.0] photometric redshift slice for this chip.
The separation between the 2 secure galaxies is only 3'', while the 4 galaxies are included in a 
40'' diameter circle, 0.44 h$_{50}^{-1}$Mpc at z$\sim$3.5 
($\sim$ 2 h$_{50}^{-1}$ Mpc for q$_0$=0.1 in comoving coordinates).

If we focus into this redshift interval, and we compute the K-S test for the added catalogue of chips 2
and 3, we find a probability P$_{KS}$=0.3\%. When considering the whole field (chips 2+3+4), 
the probability for this redshift range remains low, P$_{KS}$=0.5\%. 
We have also tested a redshift interval centered on z=3.4, with a width of 0.5 as [3.15;3.65], and 
the result found is P$_{KS}$=0.9\% for chip 2. It is worth to note that, according to the
simulations in \S2.3.2, a mean overestimate of the \zphot is expected for objects at z$\geq$2.8.
The mean value of this bias is 0.05, but it reaches $\sim$0.1 at 3.2$\leq$z$\leq$3.6.
For this reason, the strong signal coming from the redshift intervals containing z=3.5 is compatible
with galaxies being actually at z$\sim$3.4.

\begin{figure}
\centerline{\psfig{figure=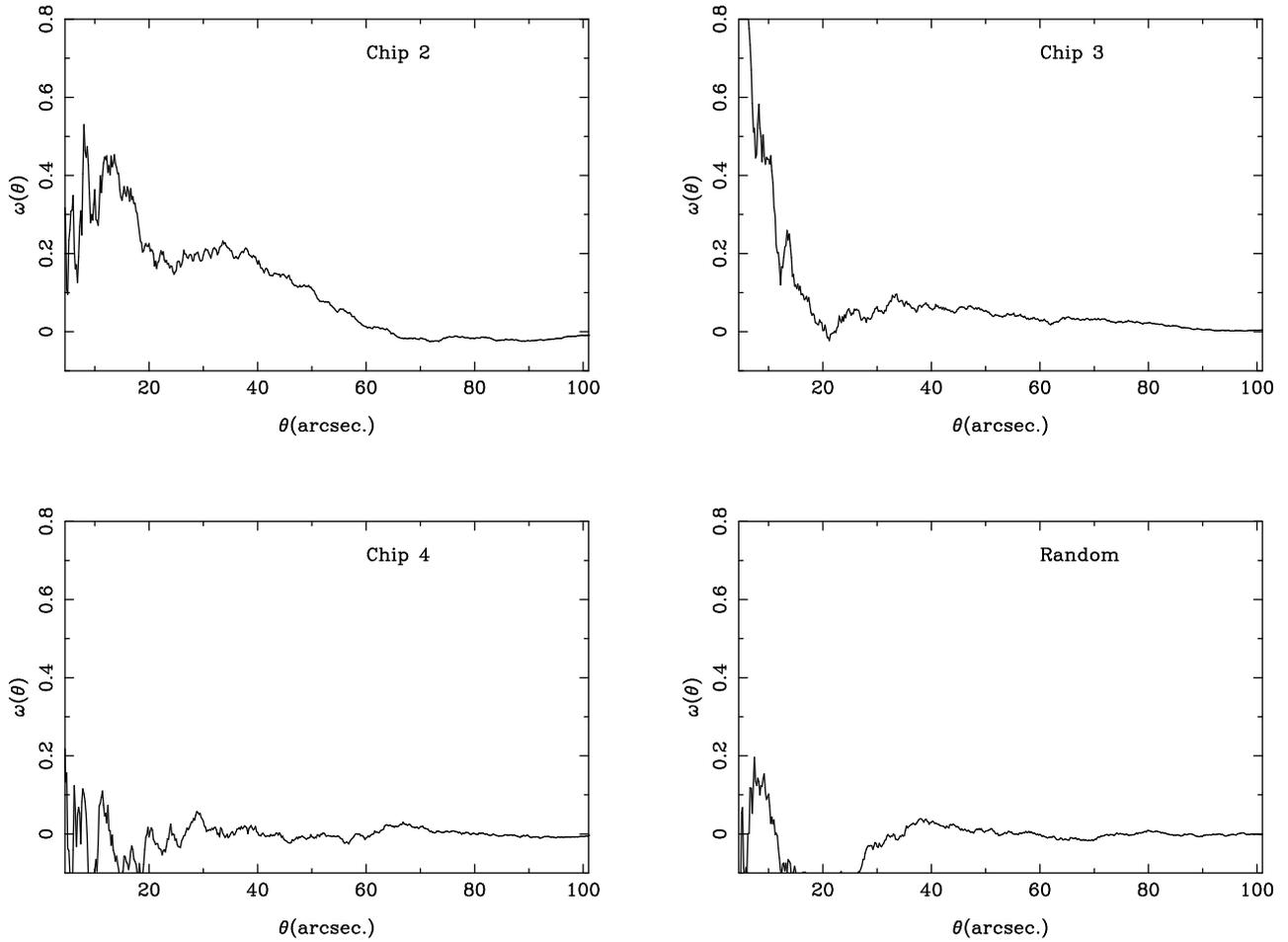,height=16cm}}
\caption{Raw correlation functions from 5'' to 100'' for galaxies included in the [3.5;4.0]
redshift interval. The results for chips 2, 3 and 4 are shown respectively on the up left,
up right and bottom left panels. The bottom right panel is the average of 100 random
catalogues with the same number of objects than in chip 2.}
\label{correl1}
\end{figure}

The K-S test gives 99.96 \% confidence on the existence of clustered galaxies at \zphot=[3.5;4.0] on 
the HDF, but it does not give information about the clustering characteristics of these galaxies.
We have examined the correlation length of these galaxies using the estimator of the correlation 
function $\omega(\theta)$  (eq. \ref{ome}) introduced by Landy \& Szalay (1993):


\begin{equation}
  \omega(\theta) = {{DD(\alpha<\theta) - 2DR(\alpha<\theta)+RR(\alpha<\theta)} \over {RR(\alpha<\theta)}}
\label{ome}
\end{equation}


where for angles $\alpha < \theta$, DD is the number of data pairs, DR is the number of pairs between 
data and random sample, and RR is the number of random pairs.
We proceed by generating a normal random sample, containing 5 times the number of real galaxies, 
distributed in the same physical area. This operation is repeated for each individual chip.
The final $\omega(\theta)$ is an average of the results for 100 different random catalogues.
At the same time we measure $\omega(\theta)$ for purely random samples, with the same field 
size and the same number of objects as the real data, to study the significance of the signal. 
We will not discuss correlations with lengths less than 10'' because our catalogue is biased 
against such scales. The raw $\omega(\theta)$ results in the redshift interval [3.5;4.0] are 
summarized in figure \ref{correl1} for the different chips, and compared to a random distribution.
We find a high correlation signal from 10'' to 50'' for galaxies on chip 2 within the redshift
interval [3.5;4.0]. 
When we compare with the other chips in the same redshift interval, only chip 3 shows
a signal on smaller scales (from 10'' to 20''). For the other redshift intervals considered in 
the K-S test, no significant signal is found with this estimator on scales larger than 20'', 
in any of the chips when computed individually. These results are consistent with the K-S test.

We have chosen to compute the number density isocontours (Dressler 1980), 
in order to unveil the structure related to these z$\sim$3.4 galaxies. 
For this purpose, we have used the whole field catalogue. Results are shown in figure \ref{iso}. 
A structure appears on chip 2, about 60'' long and 10'' wide, which corresponds to 0.66 
h$_{50}^{-1}$ Mpc $\times$ 0.11 h$_{50}^{-1}$ Mpc projected at z=3.4 for q$_0$=0.1 
(around  3 h$_{50}^{-1}$ Mpc $\times$ 0.5 h$_{50}^{-1}$ Mpc in comoving coordinates).
 It shows a main density peak and two secondary peaks along the structure suggesting 
substructures merging. It is worth noting that the strongest density peak for this 
structure is centered only $\sim$ 10'' away from the position of galaxies A and B.
The position of the structure does not change if we consider only galaxies from chip 2 
or from the whole field taken together.
We randomly remove some galaxies from our catalogues to test the stability of the position and 
the shape of the structures.
When 10 \% of the points are removed, the position of the large structure in chip 2 remains 
unchanged, and shows only a lower global intensity. When 25 \% of the objects are removed, 
the structure is still visible and the position
does not change. It is difficult to explain this structure as an artifact due to edge or 
other spurious effects. 
We can also see other structures appearing on chip 3, but their scale is smaller ($\sim$20''), 
as highlighted by the correlation test, and they are apparently disconnected. Anyway, these 
structures are less significant because their positions are not stable in front of the random 
removal of objects.

\subsection{Spatial correlation function}

The measure of the angular correlation function together with the redshift information gives a direct 
estimate of the spatial correlation function expressed in proper coordinates (Efstathiou et al. 1991):


\begin{equation}
 \xi(r,z)=\left({r_0 \over r}\right)^\gamma (1+z)^{-(3+\epsilon)}
\label{xi}
\end{equation}

\begin{figure}
\centerline{\psfig{figure=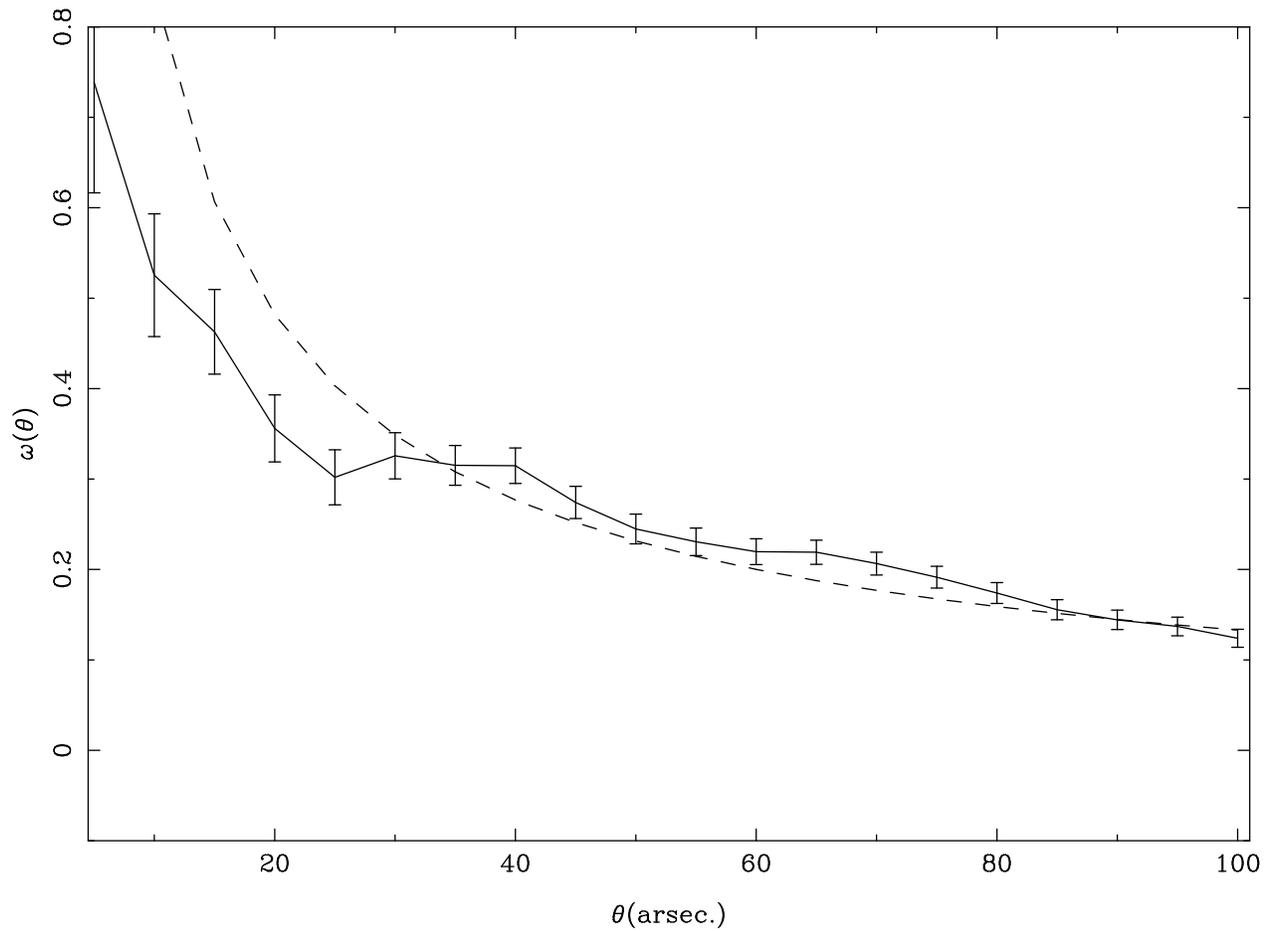,height=16cm}}
\caption{Correlation function for galaxies in the whole field, belonging to the
redshift interval [3.5;4.0]. Error bars show the 1 $\sigma$ Poisson errors. 
The dashed line is the best power law fit with a fixed $\gamma = 1.8$,
and $r_0^{\gamma} = 0.039$ (see equation \ref{wrp}).}
\label{correl2}
\end{figure}
\begin{figure}
\centerline{\psfig{figure=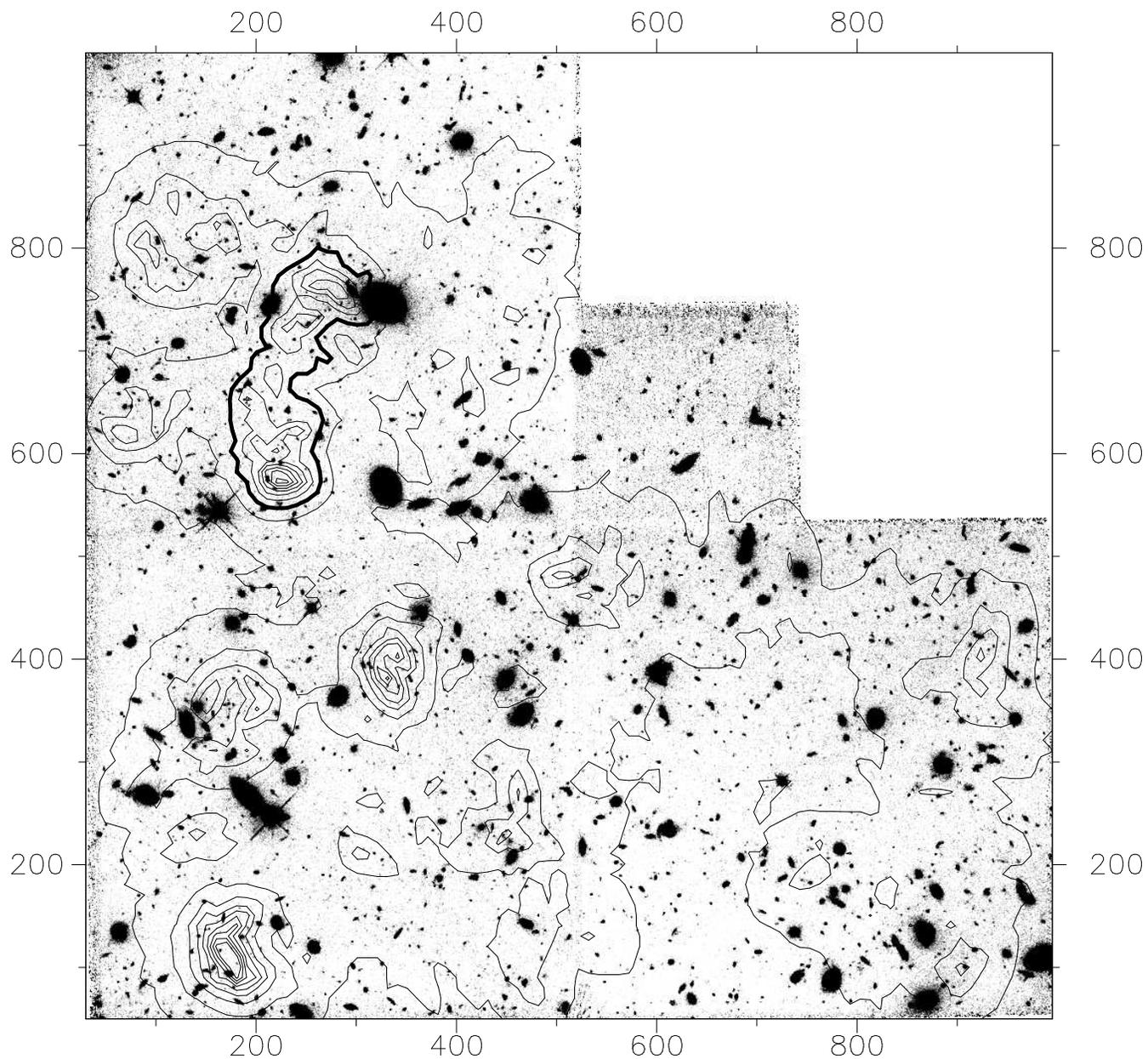,height=16cm}}
\caption{Isocontour plot of the projected number density of galaxies compatible with z$\sim$3.4, 
superimposed on the I image of HDF field. The first contour represents the mean value over the 
whole field, with successive contours increasing by 1 $\sigma$. The maximum value displayed 
is 10 $\sigma$. The thick line draws the 3 $\sigma$ contour of the main structure detected in
chip 2. Galaxy A (z=3.43) is located at (233,642) and galaxy B (z=3.37) is at (216,644).}
\label{iso}
\end{figure}

Assuming small angles, and taking into account the redshift intervals defined by the \zphot, we can 
deduce from equation \ref{xi} the real-space correlation function as the power-law (Peebles 1980, 
Carlberg et al. 1997),


\begin{equation}
\omega(r_p)= {{\Gamma(1/2)\Gamma((\gamma - 1)/2))}\over{\Gamma(\gamma/2)}}
\times r_0 ^\gamma r_p ^{(1-\gamma)}
\label{wrp}
\end{equation}


where r$_p$ is the proper separation of galaxy pairs in the projected direction, expressed in Mpc.
We choose first to fix $\gamma$=1.8, in order to compare our results to the local population of 
galaxies. The best power-law fit to the correlation function, with this $\gamma$ value, gives 
for the whole field a clustering length of $r_0 = 0.16 \pm 0.03$ h$_{50}^{-1}$ Mpc in the \zphot 
interval [3.5;4.0]. Considering a linear evolution for the clustering ($\epsilon$=0.8), and a
mean redshift of z$\sim$3.65 for the population of galaxies, the present correlation length is 
$r_0 = 4.1 \pm 0.8$ h$_{50}^{-1}$ Mpc (see fig. \ref{correl2}). For the other redshift intervals,
 the mean clustering length is $r_0 \sim 0.05$ h$_{50}^{-1}$ Mpc, which gives for the present 
correlation length 0.5$\lesssim r_0 \lesssim$2 h$_{50}^{-1}$ Mpc. The \zphot interval 
[3.0;3.5] is of special interest because it shows the strongest clustering signal after
[3.5;4.0]. We obtain a present correlation length of $r_0 \sim 3 $ h$_{50}^{-1}$ Mpc for this redshift
interval when considering the whole field. When we try to fit also the $\gamma$ value to the different
redshift intervals, the best fit is generally found for $\gamma$=1.65$\pm$0.4, but the error-bars are 
too large to conclude about this parameter. 

\section{Magnitudes, colors and star-formation rates of the clustered population at high redshift}

All the objects with \zphot observed within the [3.5;4.0] interval are U-dropouts and their
redshifts, corrected according to the simulations, actually sample the $3.4 \lesssim z \lesssim 
3.9$ domain. There are 119 galaxies of this kind on the whole HDF with R magnitudes ranging 
from 26 to 31, 85 of which are brighter than the completeness limit in magnitude (R$\leq$29.5). These
galaxies represent 13 \% of the total HDF population in our catalogue within this apparent 
magnitude range. Assuming that this population is uniformly sampling the redshift domain 
$3.4 \lesssim z \lesssim 3.9$,
their comoving density is at least $4.1 \times 10^{-3}$ h$_{50}^{3}$ $Mpc^{-3}$ with q$_0$=0.1
($1.8 \times 10^{-2}$ h$_{50}^{3}$ $Mpc^{-3}$ with q$_0$=0.5), then a factor of $\sim 50$
higher than the population of star-forming galaxies reported by ST96b at
$3.0 \lesssim z \lesssim 3.5$, with $23.5 \leq R \leq 25.0$.

\begin{figure}
\centerline{\psfig{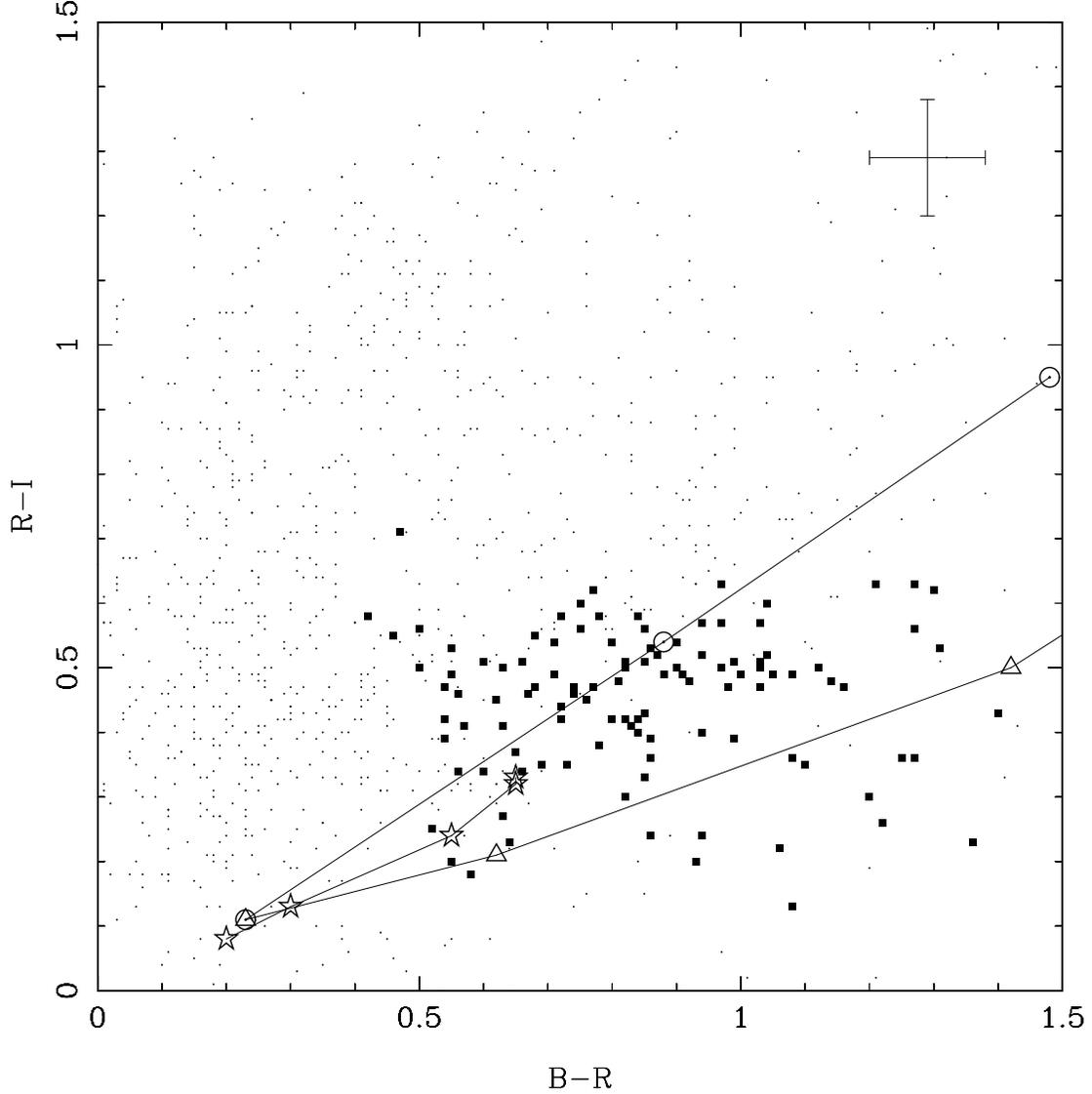}}
\caption{Color-Color BRI diagram for chip 2. Large squares correspond to the clustered 
population at $3.4 \lesssim z \lesssim 3.9$, whereas small dots are field objects. All
the high-redshift objects are U-dropouts, and their
typical error-bars are shown in the top-right corner.
The location of the solar metallicity models for the stellar population are also shown. 
Triangles correspond to the burst model with no-reddening and ages $10^6$, $10^7$ and 
$10^8$ yrs, increasing redwards. The location of the $A_V=0.0$, $A_V=0.8$ and $A_V=1.6$ 
burst models (with age $10^6$ yrs) are plotted by circles, increasing redwards. 
Stars are for the constant star-forming model without reddening, with ages $10^6$, $10^7$, 
$10^8$ and $10^9$ yrs, also increasing redwards.}
\label{bri}
\end{figure}

The color-color BRI diagram for the whole HDF distribution is plotted in figure \ref{bri}.
The 119 galaxies belonging to the sample are located in a particular region of this diagram
($B-R \sim 0.5$ to $0.9$ , and $R-I \sim 0.4$ to $0.6$). The SEDs of these
objects can be fitted by different synthetic stellar populations, and there is a degeneracy
at least in the SFR-age-metallicity-extinction space. Nevertheless, when the IMF and the
upper mass limit for star-formation are fixed, the allowed parameter space can be roughly 
constrained. We have used the GISSEL96 code for this exercise,
taking into account that these objects are all
necessarily dominated by massive stars at the wavelengths seen by the HDF.
Two kinds of SFRs were considered: a
single stellar population (instantaneous burst), and a continuous star-forming system, 
both with the Scalo IMF (1986), an upper mass-limit for stars of $125 M_{\odot}$,
and an extinction curve of SMC type given by Pr\'evot et al. (1984). 
When the burst-model is used, the observed SEDs can be fitted only by a population of
stars younger than 0.1 Gyr, with a rest-frame reddening lower than $A_V \sim 1.6$, and these
values are stable in front of metallicity changes. The best fits with a burst-model
are obtained with ages ranging from $10^6$ to $10^8$ yrs, $10^7$ yrs and $A_V \sim 0.6$
being the mean values. When the constant star-forming model is used, the observed SEDs
can be fitted with ages ranging from $10^6$ to $10^9$ yrs, and $0.3 < A_V < 0.8$, the
best age-$A_V$ fit being metallicity dependent. For simplicity, only
the locations of the solar metallicity models are given in figure \ref{bri}.

The two galaxies A and B spectroscopically identified at $z=3.4$ by LW97 have apparent magnitudes
$R=27.5 $ and $R=26.8$ respectively. Galaxy A has a rest-frame 1500 \AA\ luminosity
of $L_{1500}= 3.1 $ to $ 3.4 \times 10^{40}$ h$_{50}^{-2}$ ergs s$^{-1}$ \AA$^{-1}$ with q$_0$=0.1
($L_{1500}= 1.2 $ to $ 1.3 \times 10^{40}$ h$_{50}^{-2}$ ergs s$^{-1}$ \AA$^{-1}$ with q$_0$=0.5),
depending on details of the spectra, as given by the best fit models mentioned above, without
any correction for extinction. The mean weighted luminosity over the $26.0 \lesssim R \lesssim 29.5$
domain, assuming that galaxies are uniformly distributed over the $3.4 \lesssim z \lesssim 3.9$
interval, is $L_{1500}=  2.8 \times 10^{40}$ h$_{50}^{-2}$ ergs s$^{-1}$ \AA$^{-1}$ with q$_0$=0.1
($L_{1500}= 1.1 \times 10^{40}$ h$_{50}^{-2}$ ergs s$^{-1}$ \AA$^{-1}$ with q$_0$=0.5). The 
weighted averaged SFR obtained for this sample through a continuous star-forming model, without
any correction for extinction, is then $2.6 M_{\odot}$ h$_{50}^{-2}$ $yr^{-1}$ 
($1.0 M_{\odot}$ h$_{50}^{-2}$ $yr^{-1}$) with
q$_0$=0.1 (0.5). This value is a factor of 10 lower than the mean SFR obtained by 
ST96b, but the total star formation rate per comoving volume is about
6 times higher, $1.1 \times 10^{-2}$ h$_{50}$ $M_{\odot}$ $yr^{-1}$ $Mpc^{-3}$ 
($1.8 \times 10^{-2}$ h$_{50}$ $M_{\odot}$ $yr^{-1}$ $Mpc^{-3}$) with q$_0$=0.1(0.5).
It is worth to note that this is a lower limit, and that any correction for extinction, 
according to the best-fit models mentioned above, will tend to increase this value. 
In particular, taking the maximum reddening allowed to the continuous star-forming models, 
which implies a correction of about 1.8 magnitudes to the UV rest-frame sampled in R, we 
obtain an averaged SFR which is about 5 times higher than the precedent value.

Using the models mentioned above, and assuming that galaxies are uniformly distributed 
within the $3.4 \lesssim z \lesssim 3.9$ interval, with UV rest-frame apparent magnitudes 
$26.0 \lesssim R \lesssim 29.5$, we obtain a distribution in absolute magnitude M$_R$
ranging from -22.4 to -17.1. The mean magnitude of the sample is R=28.2, which 
corresponds to $M_R = -18.7$ to $ -20.3$ ($M_B = -18.7$ to $ -19.8$), and the widths of the
permitted intervals in magnitude are fixed by the different metallicity-age-reddening
fits to the models. The mean absolute magnitude is 0.5 to 1.5 magnitudes fainter than 
the local M$^*$ (Lin et al., 1996), depending on the models and filters, and roughly $10 \%$
of the sample is expected to be brighter than M$^*_B$. These values are extremely 
model dependent, because the wavelength range sampled
by the HDF at such high-redshifts is relatively narrow and quite sensitive to short
time-scale phenomena, making difficult to fit the observed SEDs by a synthetic stellar 
population. The near-IR photometry should be useful to constrain the parameter space. 
According to our modelling, the mean expected colors for this sample are $R-J \sim 1$ 
and $R-K \sim 2-3$. The expected IR magnitudes will be $25.0 \lesssim J \lesssim 28.5$
and $23.0 \lesssim  K \lesssim 27.5$, and $15 \%$ of the sample ($N \sim 18-20$)
should be detected with $J \lesssim 26.5$ and $K \lesssim 24.5$.

\section{Discussion and Perspectives}

We have compared the star formation density computed in \S4,
in the $3.4 \lesssim z \lesssim 3.9$ interval, to the results on the 
star formation history by Madau et al. (1996). The data points presented in
their figure 9 at high-redshift are lower limits, coming from ST96a
at $3.0 \lesssim z \lesssim 3.5$, and from a direct identification of U and B 
dropouts in the HDF at $2.5 \lesssim z \lesssim 4.5$. Surprisingly, our 
star formation density in the $3.4 \lesssim z \lesssim 3.9$ interval is similar 
to their HDF results in the lower adjacent domain ($2.5 \lesssim z \lesssim 3.5$),
and then {\it incompatible} with a global decrease of the star formation in this 
redshift interval. In addition, our integrated star formation rate
is a lower limit, not only because of completeness and reddening effects, as 
mentioned in \S4, but also because the simulations of \zphot (\S2.2) show that the 
population in this interval can be underestimated (up to $\sim 20\%$ of the
faint objects could be lost). Taking into account these effects,
except completeness, we find $log \rho_{*} = -1.7 $ ($M_{\odot}$ $yr^{-1}$ $Mpc^{-3}$)
(uncorrected) to  $log \rho_{*} = -1.0$ ($M_{\odot}$ $yr^{-1}$ $Mpc^{-3}$)
(corrected for reddening and \zphot systematics), with $H_0=50 km s^{-1} Mpc^{-1}$ 
and $q_0=0.5$, the same units and cosmology as in Madau et al. (1996). 

About $20 \%$ of the total sample detected at $3.4 \lesssim z \lesssim 3.9$
is included within the 3 $\sigma$ contour of the main structure detected in chip 2.
The evidence for such structures leads to a question: are they the progenitors of
the nowadays clusters or groups of galaxies ? In a more general way, the problem
is to interpret these high-redshift galaxies and their spatial distribution in terms of
structure formation. To answer these questions, it is important 
to compare the present results with the theoretical simulations of 
galaxy formation and high-redshift clustering.
Baugh et al. (1997), using a semi-analytic model for galaxy formation in a hierarchical
scenario, have produced a reasonable fit to the star formation data by ST96b 
and Madau et al. (1996). Their predicted star formation rate density is 
even in better agreement with the present results (see their Fig. 16). Also, the
mean SFRs for galaxies at $z \gtrsim 3$ are expected to be a few solar masses per
year, as observed in our sample. Concerning the
clustering, these simulations predict a comoving length of r$_0 \sim 4h^{-1} $Mpc,
which reproduces the population of galaxies of ST96b and the
present clusters of galaxies,
with standard CDM models, a bias parameter $b \sim 4$, and $\Omega_0$=1.0 (model A, 
$\Lambda_0$=0, $H_0$=50, b=4.2) or $\Omega_0$=0.3 (model G, $\Lambda_0$=0.7 and $H_0$=60, 
b=3.5). The present clustering length is also in good agreement with this value,
as well as with the typical length for IRAS galaxies (Fisher et al., 1994), and
with the clustering at low redshifts (Loveday et al. 1995). It is also fully compatible
with the previous correlation length measured by Villumsen et al. (1997). The 
conclusions are similar when we compare with the high-redshift cluster modelling by 
Moscardini et al. (1997). 

According to the hierarchical scenario, a structure such as the main one detected in chip
2 is probably the progenitor of a group or a cluster of galaxies. Its shape,
presenting several substructures, is also what we should expect in a hierarchical model
of structure formation (Huss, Jain \& Steinmetz 1997). In general, the structures
observed at $3.4 \lesssim z \lesssim 3.9$ are expected to be the progenitors of present-day
groups or clusters of galaxies. The present results
are consistent with a linear evolution regime for the clustering
since z $\sim$ 3.4, with $\epsilon \sim$ 0.8. 

An important result is that we observe the strongest clustering on large scales in the 
$3.4 \lesssim z \lesssim 3.9$ interval, and also a clear signal on similar scales in the
$2.9 \lesssim z \lesssim 3.4$ interval (\zphot interval [3.0;3.5]). In both cases,
the signal could be associated, at least in part, to a population of galaxies at 
z$\sim$3.4, the spectroscopic redshift of 2 objects belonging to the main structure 
. The clustering signal reduces when we consider 
the population at lower or at higher redshifts, but it is still clearly present at 
small scales, at least in chips 2 and 3. Taking into account the results
by Steidel at al. (1998) in the $2.0 \lesssim z \lesssim 3.4$ interval, the low 
signal detected in this interval compared to z$\sim$3.4 could be due to
an effect of the reduced size of the present field. At z$\gtrsim$4.0, the main 
problems are the completeness of the sample and the accuracy of \zphot, which is
dominated by photometric errors when the objects become extremely faint.

We cannot rule out completely that the strong signal detected at z$\sim$3.4 could
be due only to a projection effect on different high-redshift planes. Besides, the 
combined effect of field size and completeness in the different filters
 tend to favour the detection of a particular length scale at a given
redshift, and then the relative strength of the clustering in the different redshift
intervals has to be taken with caution. In any case, we cannot generalize the
present results based on a single deep field, and further investigation 
in other different and deep regions is required to confirm them, combined if 
possible with an extended spectroscopic survey. Photometry in the near-IR would be
useful to improve the modelling of the SEDs and the star formation estimate
(see Connolly et al. 1997). But this exercise is difficult with the present 
ground-based instruments, taking into account the expected magnitudes of the sources (\S4) 
and their typical sizes, which require an excellent spatial resolution. 
About $10-20 \%$ of the whole sample should be detected with the NICMOS images of HDF.
Concerning the main structure in chip 2, radio observations aiming to
detect the Sunyaev-Zel'dovich effect (Sunyaev \& Zel'dovich, 1980), 
along the line of sight of the HDF, could also help to confirm the existence of 
a massive structure. In a more prospective way, a weak lensing analysis in 
this field could be greatly improved by the prior knowledge of the mass distribution 
derived from the light (Bonnet \& Mellier 1995, Kaiser et al. 1995, 
Van Waerbeke et al. 1997), as given by the present study. 
It could be possible then to estimate the total mass of these structures
at high redshift and to constrain $\Omega$.

\begin{acknowledgements}

We would like to thank G. Mathez, J.P. Kneib, G. Bruzual and D. Valls-Gabaud for multiple fruitful 
discussions and advices. J. Bezecourt, J.F. Le Borgne, Y. Mellier and L. Van Waerbeke have made a careful 
reading of the paper, and their interesting comments are especially acknowledged.
Thanks also to Bob Williams and the whole HDF team to have provided these superb images. JMM is 
very grateful to the Rotary Club d'Andorra for financial support. Part of this work has been
supported by an A.D.I. Research Grant and by the Ministeri d'Educaci\'o del Govern d'Andorra. 

\end{acknowledgements}

{}

\begin{deluxetable}{lccc}
\tablecaption{Photometric properties of the catalogue}
\tablehead{
\colhead{Filter} & \colhead{completeness} & \colhead{limiting} & 
\colhead{$\mu_{\lambda} (1 \sigma$, 1 pixel)} \nl
  & \colhead{magnitude} & \colhead{magnitude} & \colhead{mag/"$^2$} \nl}
\startdata
U (F300W) & 28.0  &  29.5  &   27.00 \nl
B (F450W) & 30.0  &  32.0  &   29.00 \nl
R (F606W) & 29.5  &  32.0  &   29.00 \nl
I (F814W) & 29.0  &  31.0  &   28.50 \nl
\enddata 
\label{cat}
\end{deluxetable}

\begin{deluxetable}{lcccccc}
\tablecaption{Mean photometric errors derived through SExtractor}
\tablehead{ &  & & \colhead{magnitudes} & & & \nl 
\colhead{Filter}& \colhead{20-25}& \colhead{25-26}& \colhead{26-27} &
\colhead{27-28}& \colhead{28-29}& \colhead{ $>$ 29} \nl}
\startdata
U (F300W) & 0.03  & 0.07 & 0.13 & 0.21 & 0.30 & 0.32 \nl
B (F450W) & 0.003 & 0.01 & 0.02 & 0.05 & 0.10 & 0.23 \nl
R (F606W) & 0.003 & 0.01 & 0.02 & 0.03 & 0.07 & 0.20 \nl
I (F814W) & 0.005 & 0.02 & 0.03 & 0.07 & 0.15 & 0.25 \nl
\enddata
\label{err1}
\end{deluxetable}

\begin{deluxetable}{llccccccc}
\tablecaption{Dispersion in the photometric redshift as a function of the
photometric errors and galaxy types. For each simulated sample, four 
different informations are given: the standard deviation ($\sigma z$ (z)),
the mean systematic bias ($\Delta z = z(model) - $ \zphot), 
the mean individual uncertainties at 75 \% confidence level ($\Delta z (75 \%)$),
and the percentage of catastrophic identifications ($c\%$). See text 
for more details.}
\tablehead{ & & &  & & & & &  \nl
\colhead{Galaxy type} &  \colhead{$\Delta m$}  &  & 
\colhead{0-0.7} & \colhead{0.7-1.8} & \colhead{1.8-2.8} & \colhead{2.8-3.4}
 & \colhead{3.4-4.5} & \colhead{4.5-5.0} \nl}
\startdata
E/S0 & $\le 0.1$  & $\sigma z$ & 0.10 & 0.17 & 0.12 & 0.08 & 0.19 & 0.05  \nl
     &             & $\Delta z$ &-0.07 & 0.03 & 0.02 & -0.03 & -0.02 & -0.13  \nl
     &             & $\Delta z(75 \%)$& 0.14 & 0.22 & 0.16 & 0.05 & 0.26 & 0.05  \nl
     &             & $c\%$ & 4 & 2 & 7 & $<$1 & 5 & $<$1  \nl
C. SFR & $\le 0.1$ & $\sigma z$ & 0.07 & 0.25 & 0.10 & 0.06 & 0.12 & 0.04  \nl
     &             & $\Delta z$ &-0.03 & 0.04 & 0.02 & -0.07 & -0.06 & -0.14  \nl
     &             & $\Delta z(75 \%)$& 0.15 & 0.36 & 0.17 & 0.04 & 0.24 & 0.06  \nl
     &             & $c\%$ & 4 & 10 & 11 & $<$1 & 11 & 2 \nl
all  & $\le 0.1$   & $\sigma z$ & 0.08 & 0.22 & 0.11 & 0.07 & 0.15 & 0.04 \nl
     &             & $\Delta z$ &-0.03 & 0.03 & 0.00 &-0.05 &-0.05 &-0.14 \nl
     &             & $\Delta z(75 \%)$& 0.14 & 0.30 & 0.16 & 0.04 & 0.25 & 0.05 \nl
     &             & $c\%$ & 4 & 10 & 2 & $<$1 & 19 & $<$1  \nl
all  & 0.2  & $\sigma z$ & 0.10 & 0.37 & 0.15 & 0.10 & 0.21 & 0.06 \nl
     &      & $\Delta z$ &-0.04 & 0.05 &-0.03 &-0.06 &-0.11 & -0.13 \nl
     &      & $\Delta z(75 \%)$& 0.27 & 0.56 & 0.31 & 0.11 & 0.33 & 0.09  \nl
     &      & $c\%$ & 12 & 12 & 7 & $<$1 & 38 & $<$1  \nl
all  & 0.3  & $\sigma z$ & 0.23 & 0.44 & 0.20 & 0.17 & 0.24 & 0.09 \nl
     &      & $\Delta z$ &-0.07 & 0.08 &-0.03 &-0.03 &-0.14 & -0.13 \nl
     &      & $\Delta z(75 \%)$& 0.35 & 0.57 & 0.40 & 0.17 & 0.39 & 0.12  \nl
     &      & $c\%$ & 23 & 12 & 11 & $<$1 & 49 & $<$1  \nl
\enddata
\label{sim1}
\end{deluxetable}

\begin{deluxetable}{cccc}
\tablecaption{Kolmogorov-Smirnov Test Results}
\tablehead{\colhead{Chip Number} & \colhead{Redshift Interval} & 
\colhead{No. of Galaxies} & \colhead{P$_{KS} \% $} \nl}
\startdata
  2 & 2.5--3.0 & 53 & 17.0 \nl
    & 3.0--3.5 & 47 & 2.5 \nl
    & 3.5--4.0 & 43 & 0.4 \nl
    & 4.0--4.5 & 25 & 19.0 \nl
  3 & 2.5--3.0 & 48 & 3.5 \nl
    & 3.0--3.5 & 54 & 6.2 \nl
    & 3.5--4.0 & 44 & 3.9 \nl
    & 4.0--4.5 & 14 & 20.8 \nl
  4 & 2.5--3.0 & 50 & 44.9 \nl
    & 3.0--3.5 & 38 & 50.8 \nl
    & 3.5--4.0 & 32 & 7.1 \nl
    & 4.0--4.5 & 15 & 68.8 \nl
\enddata
\label{ks}
\end{deluxetable}

\begin{thebibliography}{}

\bibitem{}Baugh, C.M., Cole, S., Frenk, C.S., Lacey, C.G., 1997,
 \apj, submitted, astro-ph/9703111

\bibitem{}Bertin, E., Arnouts, S., 1996, \aaps, 117, 393

\bibitem{}Bonnet, H., Mellier, Y., 1995, \aap, 303, 331

\bibitem{}Burrows, C.J., Bagget, S.M., Biretta, J., 1995, WFPC2 Instrument 
Handbook V3.0, C.J. Burrows (Ed.), STScI publication

\bibitem{}Brunner, R.J., Connolly, A.J., Szalay, A.S., Bershady,
 M.A., 1997, \apjl, 428, 21

\bibitem{}Bruzual, G., Charlot, S., 1993, \apj, 405, 538

\bibitem{}Bruzual, G., Charlot, S., 1997, \apj, in preparation

\bibitem{}Calberg, R.G., Cowie, L.L., Songaila, A., Hu, E.M., 1997,
 \apj, 484, 538

\bibitem{}Cohen, J.G., Cowie, L.L., Hogg, D.W., Songaila, A., Blanford,
 R., Hu, E.M., Shopbell, P., 1996, \apjl, 471, 5

\bibitem{}Cowie, L.L., 1997, http://www.ifa.hawaii.edu/cowie/tts/tts.html

\bibitem{}Connolly, A.J., Csabai, I., Szalay, A.S., Koo, D.C., Kron, R.G.,
 Munn, J.A., 1995, \aj, 110, 6

\bibitem{}Connolly, A.J. et al., 1996, \apjl, 473, 67

\bibitem{}Connolly, A.J., Szalay, A.S, Dickinson, M., SubbaRao, M.,
 Brunner, R.J., 1997, \apjl, 486, 11

\bibitem{}Dressler, A., 1980, \apjs, 42, 565

\bibitem{}Dressler, A., Oemler, A., Gunn, J.E., Butcher, H., 1993,
 \apjl, 404, 45

\bibitem{}Efstathiou, G., Bernstein, G., Katz, N., Tyson, J.A,
 Guhathakurta, P., 1991, \apjl, 380, 47

\bibitem{}Fisher, K.B., Davis, M., Strauss, M.A., Yahil, A., Huchra,
 J.P., 1994, \mnras, 267, 927

\bibitem{}Francis, P.J. et al., 1996, \apj, 457, 490

\bibitem{}Giavalisco, M., Steidel, C.C., Szalay, A.S., 1994, \apj,
 425, 5

\bibitem{}Gwyn, S., Hartwick, F., 1996, \apjl, 468, 77

\bibitem{}Huss, A., Jain, B., Steinmetz, M., 1997, \mnras, submitted,
 astro-ph/9703014

\bibitem{}Kaiser, N., Squires, G., Broadhurst, T.J., 1995, \apj,
 449, 460

\bibitem{}Landy, S.D., Szalay, A.S., 1993, \apjl, 412, 64

\bibitem{}Lanzetta, K.M., Yahil, A., Fernandez-Soto, A., 1996,
 \nat, 381, 759 (LYF)

\bibitem{}Le F\`evre, O., Deltorn, J.M., Crampton, D., Dickinson,
 M., 1996, \apjl, 471, 11

\bibitem{}Lin, H., Kirshner, R.P., Schetman, S.A., et al., 1996,
 \apj, 464, 60

\bibitem{}Loveday, J., Maddox, S.J., Efstathiou, G., Peterson, B.A.,
 1995, \apj, 442, 457

\bibitem{}Lowenthal, J.D., Koo, D.C., Guzman, R., et al., 1997, \apj,
 481, 673 (LW97)

\bibitem{}Madau, P., Ferguson, H.C., Dickinson, M.E., Giavalisco, M.,
 Steidel, C.C., Fruchter, A., 1996, \mnras, 283, 1388

\bibitem{}Moscardini, L., Coles, P., Luchin, F., Matarrese, S., 1997,
 \mnras, submitted, astro-ph/9712184 

\bibitem{}Peacock, J.A., 1983, \mnras, 202, 615

\bibitem{}Peebles, P.J.E., 1980, The Large-Scale Structure of the
 Universe (Princeton: Princeton Univ. Press)

\bibitem{}Pell\'o, R., Miralles, J.-M., Le Borgne, J.-F., Picat, J.-P.,
 Soucail, G., Bruzual, G., 1996, \aap, 314, 73

\bibitem{}Phillips, S., 1985, \mnras, 212, 657

\bibitem{}Pr\'evot M.L., Lequeux , J., Pr\'evot L., Maurice, E.,
 Rocca-Volmerange B., 1984, \aap, 132, 389

\bibitem{}Sawicki, M.J., Lin, H., Yee, H.K.C., 1997, \aj, 113,
 1 (SLY)

\bibitem{}Scalo J.M., 1986, \fcp, 11, 1

\bibitem{}Steidel, C.C., Giavalisco M., Dickinson, M., Adelberger,
 K.L., 1996a, \aj, 112, 352 (ST96a)

\bibitem{}Steidel, C.C., Giavalisco M., Pettini, M., Dickinson, M.,
 Adelberger, K.L., 1996b, \apj, 462, L17 (ST96b)

\bibitem{}Steidel, C.C., Adelberger, K.L., Dickinson, M., Giavalisco M.,
  Pettini, M., Kellog, M., 1998, \apj, 492, 428 

\bibitem{}Subbarao M.U., Connolly A.J., Szalay A.S., Koo D.C.,
 1996, \aj, 112, 929

\bibitem{}Sunyaev, R.A., Zel'dovich, Y.B., 1980, \araa, 18, 537

\bibitem{}Van Waerbeke, L., Mellier, Y., Schneider, P., Fort, B.,
 Mathez, G., 1997, \aap, 317, 303

\bibitem{}Villumsen, J.V., Freudling, W., Da Costa, L.N., 1997,
 \apj, 481, 578 

\bibitem{}Williams, R.E., Blacker, B., Dickinson, M., et al., 1996, \aj, 112, 1335

\bibitem{}White, S.D.M, 1996, astro-ph/9608044

\bibitem{}Zepf, S.E., Moustakas, L.A., Davis, M. , 1997, \apjl,
 474, 1

\end{thebibliography}
\end{document}